\documentclass[a4paper,12pt]{article}
\usepackage[english]{babel}
\usepackage{amsmath}
\usepackage{graphicx}
\usepackage[utf8]{inputenc}
\usepackage{amssymb}
\usepackage{cite}
\usepackage{amsfonts}
\usepackage[colorlinks=true,urlcolor=blue,citecolor=blue]{hyperref}
\usepackage{url}
\usepackage{doi}
\providecommand{\U}[1]{\protect\rule{.1in}{.1in}}
\providecommand{\U}[1]{\protect\rule{.1in}{.1in}}
\begin{document}
	
	\title{Exact solutions for time-dependent complex \\
		symmetric potential well }
	\author{{\small B. Khantoul}$^{a,b}$\thanks{%
			E-mail: boubakeur.khantoul@univ-constantine3.dz} \quad {\small and} \quad 
		{\small A. Bounames}$^{a}$\thanks{%
			E-mail: bounames@univ-jijel.dz} \\
		%EndAName
		$^{(a)}${\small Laboratory of Theoretical Physics,} {\small Department of
			Physics, University of Jijel,}\\
		\ \ \ \ {\small BP 98 \ Ouled Aissa, 18000 Jijel, Algeria.}\\
		$^{(b)}${\small Department of Process Engineering, University of Constantine
			3 - Salah Boubnider,}\\
		\ \ \ {\small BP \ B72 Ali Mendjeli, 25000 Constantine, Algeria.}}
	\date{}
	\maketitle

	\begin{abstract}
		Using the pseudo-invariant operator method, we investigate the model of a
		particle with a time-dependent mass in a complex time-dependent symmetric
		potential well $V\left( x,t\right) =if\left(t\right) \left\vert x\right\vert$. 
		The problem is exactly solvable and the analytic expressions of the
		Schr\"{o}dinger wavefunctions are given in terms of the Airy function. Indeed,
		with an appropriate choice of the time-dependent metric operators and the
		unitary transformations, for each region, the two corresponding
		pseudo-Hermitian invariants transform into a well-known time-independent
		Hermitian invariant which is the Hamiltonian of a particle confined in a
		symmetric linear potential well. The eigenfunctions of the last invariant are
		the Airy functions. Then, the phases obtained are real for both regions and the
		general solution to the problem is deduced.\\\\

\end{abstract}
	%\maketitle
	
\section{Introduction}

The discovery of a class of non-Hermitian Hamiltonian that may have a real
spectrum has prompted a revival of theoretical and applied research in quantum
physics. In fact, in 1998, C.M. Bender and S. Boettcher showed that any
non-Hermitian Hamiltonian invariant under the unbroken space-time reflection,
or $\mathcal{PT}$-symmetry, has real eigenvalues and satisfies all the
physical axioms of quantum mechanics \cite{Bender1,ZnojilL,BenderL}. In 2002,
A. Mostafazadeh presented a more extended version of non-hermitian
Hamiltonians having a real spectrum, proving that the hermiticity of the
Hamiltonian with respect to a positive definite inner product, 
$\left\langle.,.\right\rangle _{{\eta}}=\left\langle.\right\vert{\eta}\left\vert.\right\rangle$, is a necessary and sufficient condition for the reality of
the spectrum, where $\eta$ is the metric operator which is linear, hermitian,
invertible and positive. This condition requires that the Hamiltonian $H$
satisfies the pseudo-Hermitian relation \cite{Mosta1,Mosta2,Mosta3}
\begin{equation}
H^{\dagger}={\eta}H{\eta^{\dagger}.}\label{1}%
\end{equation}
Moreover in recent years, a significant progress has been achieved in
the study of time-dependent (TD) non-hermitian quantum systems in several
branches of physics. Finding exact solutions to the TD Schr\"{o}dinger equation,
which cannot be reduced to eigenvalues equation in general, is a problem of
intriguing difficulty. Different methods are used to obtain solutions of
Schr\"{o}dinger's equation for explicitly TD systems, such as unitary and
non-unitary transformations, the pseudo-invariant method, Dyson's maps, point
transformations, Darboux transformations, perturbation theory and adiabatic
approximation 
\cite{Choutri,Yuce,Sousa,Faria1,Mosta4,Znojil1,Znojil2,Wang1,Fring1,Fring2,Khantoul,Maamache1,Maamache2,Baghchi,Ramos,Koussa1,Wang,Cheniti,Znojil4,Cherbal,Choi,Luiz,Mosta5,Moussa2,Gu}.

However, the emergence of a non-linear Ermakov-type auxiliary equation for several TD systems, which is difficult to solve, constitutes an additional constraint to obtain exact analytical solutions \cite{Ermakov,Schuch}. This greatly
reduces the number of exactly solvable time-dependent non-hermitian systems
\cite{Frith,Koussa2,Tenney,Zelaya,Wu}. In particular, other works have been concerned with studying exact solutions of TD Hamiltonians with a specific TD mass in the non-Hermitian case \cite{Kecita,Martinez2} and also in the Hermitian case \cite{caldirola,kanai,abdalla,ramos2,Zelaya2}.
  
In the present work, we used the pseudo-invariant method \cite{Khantoul} to
obtain the exact solutions of the Schr\"{o}dinger equation for a particle with TD
mass moving in a TD complex symmetric potential well
\begin{equation}
V\left( x,t\right)  =if\left(  t\right)  \left\vert x\right\vert ,\label{pot}%
\end{equation}
where $f(t)$ is an arbitrary real TD function.\\
The manuscript is organised as follows: In section 2, we introduce some of the
basic equations of the TD non-hermitian Hamiltonians and their time-dependent
Schr\"{o}dinger equation (TDSE) with a TD metric. In section 3, we discuss the use
of the Lewis-Riesenfeld invariant method to address the Schr\"{o}dinger equation
for an explicitly TD non-hermitian Hamiltonian. In section 4, we use the
Lewis-Riesenfeld method to solve the TD Schr\"{o}dinger equation for a particle
with TD mass in a TD complex symmetric potential well. Finally, in Section 5,
we conclude with a brief review of the obtained results.

\section{TD Non-hermitian Hamiltonian with TD metric}

Let $H(t)$ be a non-Hermitian TD Hamiltonian and $h\left(  t\right)  $ its
associated TD hermitian Hamiltonian$.$ The two corresponding TD
Schr\"{o}dinger equations describing the quantum evolution are
\begin{equation}
H(t)\left\vert \Phi^{H}(t)\right\rangle =i\hbar\partial_{t}\left\vert \Phi
^{H}(t)\right\rangle ,\label{shrod2}%
\end{equation}%
\begin{equation}
h\left(  t\right)  \left\vert \Psi^{h}(t)\right\rangle =i\hbar\partial
_{t}\left\vert \Psi^{h}(t)\right\rangle ,\label{PSCH}%
\end{equation}
where the two Hamiltonians are related by the Dyson maps $\rho\left(t\right)$ as
\begin{equation}
H\left(  t\right)  =\rho^{-1}\left(  t\right)  h\left(  t\right)  \rho\left(
t\right)  -i\hbar\rho^{-1}\left(  t\right)  \dot{\rho}\left(  t\right)
,\label{quasi}%
\end{equation}
and their wavefunctions $\left\vert \Phi^{H}(t)\right\rangle $ and $\left\vert
\Psi^{h}(t)\right\rangle $ as
\begin{equation}
\left\vert \Psi^{h}(t)\right\rangle =\rho\left(  t\right)  \left\vert \Phi
^{H}(t)\right\rangle .\label{vect}%
\end{equation}

The hermiticity of $h(t)$ allowed us to establish the connection between the
Hamiltonian $H\left(  t\right)  $ and its Hermitian conjugate $H^{\dag}(t)$
as\textit{\ }%
\begin{equation}
H^{\dag}\left(  t\right)  =\eta\left(  t\right)  H\left(  t\right)  \eta
^{-1}\left(  t\right)  +i\hbar\dot{\eta}\left(  t\right)  \eta^{-1}\left(
t\right)  ,\label{PHH1}%
\end{equation}
which is a generalisation of the well-known conventional quasi-Hermiticity relation $\left( \ref{1}\right)$, and the 
TD metric operator is hermitian and defined as $\eta(t)=\rho^{\dag}\left(  t\right)  \rho\left(t\right)$.

\section{Pseudo-invariant operator method}

Let us start with the description of the Lewis-Riesenfeld theory \cite{Lewis}
for a TD Hermitian Hamiltonian $h\left(  t\right)  $ with a hermitian TD
invariant $I^{h}\left( t\right) $. The dynamic invariant $I^{h}\left(
t\right) $ satisfies

\begin{equation}
\frac{dI^{h}(t)}{dt}=\frac{\partial I^{h}(t)}{\partial t}-\frac{i}{\hbar
}\left[  I^{h}\left(  t\right)  ,h\left(  t\right)  \right]  =0.  \label{LR}%
\end{equation}

The eigenvalue equation for $I^{h}\left(  t\right)  $ is
\begin{equation}
I^{h}\left(  t\right)  \left\vert \psi_{n}^{h}(t)\right\rangle =\lambda
_{n}\left\vert \psi_{n}^{h}(t)\right\rangle ,\label{Eveq}%
\end{equation}
where the eigenvalues $\lambda_{n}$ of $I^{h}\left(  t\right)  $ are reals and
time-independent, and the Lewis-Riesenfeld phase is defined as%
\begin{equation}
\hbar\frac{d}{dt}\varepsilon_{n}(t)=\left\langle \psi_{n}^{h}(t)\right\vert
i\hbar\frac{\partial}{\partial t}-h\left(  t\right)  \left\vert \psi_{n}%
^{h}(t)\right\rangle .\label{pha}%
\end{equation}
and the solution of the TDSE of $h\left(  t\right)  $ is given as%
\begin{equation}
\left\vert \Psi^{h}(t)\right\rangle =\exp\left[  i\varepsilon_{n}(t)\right]
\left\vert \psi_{n}^{h}(t)\right\rangle .
\end{equation}
In the paper \cite{Khantoul}, we showed that any TD Hamiltonian
$H(t)$ satisfying the TD quasi-hermiticity relation $\left(\ref{PHH1}\right) $ admits a pseudo-hermitician invariant $I^{ph}(t)$ such that
\begin{equation}
I^{ph\dag}\left(t\right)=\eta(t)I^{ph}\left(t\right) \eta^{-1}(t) \Leftrightarrow   
I^{h}(t)=\rho(t)I^{ph}(t)\rho^{-1}(t)=I^{h\dag}(t). \label{quas}%
\end{equation}
Since the hermitian invariant $I^{h}(t)$ satisfies the eigenvalues equation
$\left( \ref{Eveq}\right) $, Eq. $\left(\ref{quas}\right) $ ensures that the
pseudo-hermitian invariant's spectrum is real with the same eigenvalues
$\lambda_{n}$ of $I^{h}(t)$
\begin{equation}
I^{h}(t)\left\vert \psi_{n}^{h}(t)\right\rangle =\lambda_{n}\left\vert
\psi_{n}^{h}(t)\right\rangle ,
\end{equation}
\begin{equation}
I^{ph}\left(  t\right)  \left\vert
\phi_{n}^{ph}(t)_{n}(t)\right\rangle =\lambda_{n}\left\vert \phi_{n}%
^{ph}(t)\right\rangle ,\label{Iph}%
\end{equation}
where the eigenfunctions $\left\vert \psi_{n}^{h}(t)\right\rangle $ and
$\left\vert \phi_{n}^{ph}(t)\right\rangle ,$ of $I^{h}(t)$\ and $I^{ph}\left(
t\right)$, respectively, are related as%
\begin{equation}
\left\vert \psi_{n}^{h}(t)\right\rangle =\rho(t)\left\vert \phi_{n}%
^{ph}(t)\right\rangle .\text{ }%
\end{equation}
The inner products of the eigenfunctions associated with the non-Hermitian
invariant $I^{ph}(t)$ can now be written as
\begin{equation}
\langle\phi_{m}^{ph}(t)\left\vert \phi_{n}^{ph}(t)\right\rangle _{\eta
}=\langle\phi_{m}^{ph}(t)|\eta\left\vert \phi_{n}^{ph}(t)\right\rangle
=\delta_{mn},\label{11}%
\end{equation}
and it corresponds to the conventional inner product associated to the
Hermitian invariant $I^{h}(t)$.

It is easy to verify, by a direct substitution of the hermitian Hamiltonian
$h(t)$ and the hermitian invariant $I^{h}(t)$ by their equivalents in the
relations $\left(\ref{quasi}\right)  $\ and $\left(\ref{quas}\right) $, respectively, 
that the pseudo hermitian invariant $I^{ph}(t)$ satisfies
\begin{equation}
\frac{\partial I^{ph}(t)}{\partial t}=\frac{i}{\hbar}\left[  I^{ph}%
(t),H(t)\right]  .\label{NINV}%
\end{equation}

We should remark that the invariant operator's eigenstates and eigenvalues can
be computed using the same procedure as the hermitian case.

The solution $\left\vert \Phi^{H}(t)\right\rangle $ of the Schr\"{o}dinger
equation $\left(\ref{shrod2}\right)$ is different from $\left\vert
\phi_{n}^{ph}(t)\right\rangle $ in Eq. $\left(\ref{Iph}\right)$ only by
the factor $e^{i\varepsilon_{n}^{ph}(t)}$ where $\varepsilon_{n}^{ph}(t)$ is a
real phase given by%

\begin{equation}
\hbar\frac{d}{dt}\varepsilon_{n}^{ph}(t)=\left\langle \phi_{n}^{ph}%
(t)\right\vert \eta(t)\left[  i\hbar\frac{\partial}{\partial t}-H\left(
t\right)  \right]  \left\vert \phi_{n}^{ph}(t)\right\rangle .\label{phase1}%
\end{equation}

\section{Particle in TD complex symmetric potential well}
Let us consider a particle with a TD mass $m(t)$ in the presence of a pure imaginary TD symmetric potential well $\left(  \ref{pot}\right)  $, where its Hamiltonian can be written as
\begin{equation}
H\left(  t\right)  =\left\{
\begin{array}
[c]{c}%
\frac{p^{2}}{2m\left(  t\right)  }+if(t) x\text{ \ \ \ if }x\geq0\\
\frac{p^{2}}{2m\left(  t\right)  }-if(t)x\text{ \ \ \ if \ }x\leq0
\end{array}
\right.  ,\label{100'}%
\end{equation}
the associated TDSE of the system is
\begin{equation}
\left[  \frac{p^{2}}{2m\left(  t\right)  }+if(t)\left\vert x\right\vert
\right]  \Psi\left(  x,t\right)  =i\frac{\partial}{\partial t}\Psi\left(
x,t\right)  ,\label{100}%
\end{equation}
where $m\left(  t\right)  $ is the particle TD mass and $f(t)$ an arbitrary
real TD function, and the unit of $\hbar=1.$
This model can be considered as the complex version of the hermitian
case of a particle, with TD mass and charge $q$, moving under the action of TD electric field  $E\left( t\right)$ and confined in a pure imaginary symmetric linear potential well: $if\left(  t\right)  x$ for $x\geq0$ and $-if\left(  t\right)  x$ for $x\leq0$,  where $f\left(  t\right) =-qE\left(t\right)$.\newline

According to the results in Ref. \cite{Khantoul}, the solution to the TD
Schr\"{o}dinger equation with a TD non-hermitian Hamiltonian is easily found
if a nontrivial TD pseudo-Hermitian invariant $I^{ph}(t)$ exists and satisfies
the von-Neumann equation $\left(  \ref{NINV}\right)  .$

In the current problem, in order to solve the TD Shr\"{o}dinger
equation\ $\left(  \ref{100}\right)  $ we assume that the Hamiltonian $H(t)$
admits an invariant in each region: let $I_{1}^{ph}\left(  t\right)  $ for
$x\geq0$ and $I_{2}^{ph}\left(  t\right)  $ for $x\leq0$.

For the region $x\geq0$, let us look for a non-Hermitian TD invariant in the
following quadratic form
\begin{equation}
I_{1}^{ph}\left(  t\right)  =\beta_{1}\left(  t\right)  p^{2}+\beta_{2}\left(
t\right)  x+\beta_{3}\left(  t\right)  p+\beta_{4}\left(  t\right)
,\label{pI1}%
\end{equation}
where $\beta_{i}\left(  t\right)  $ are arbitrary complex functions to be
determined. By inserting the expressions $(\ref{100'})$ and $(\ref{pI1})$ in
Eq. $\left(  \ref{NINV}\right)  $, the following system of equations can be
found%
\begin{equation}
\left\{
\begin{array}
[c]{c}%
\text{ }\dot{\beta}_{1}\left(  t\right)  =0,\text{
\ \ \ \ \ \ \ \ \ \ \ \ \ \ \ \ \ \ \ \ \ \ \ \ \ \ \ \ }\\
\text{ }\dot{\beta}_{2}\left(  t\right)  =0,\text{
\ \ \ \ \ \ \ \ \ \ \ \ \ \ \ \ \ \ \ \ \ \ \ \ \ \ \ \ }\\
\dot{\beta}_{3}\left(  t\right)  =-\frac{\beta_{2}(t)}{m\left(  t\right)
}+2if\left(  t\right)  \beta_{1}(t),\text{ }\\
\text{ }\dot{\beta}_{4}\left(  t\right)  =if\left(  t\right)  \beta
_{3}(t),\text{ \ \ \ \ \ \ \ \ \ \ \ \ \ \ \ \ \ }%
\end{array}
\right. \label{sys1}%
\end{equation}
to simplify the calculations, we take $\beta_{1}(t)=1$ and $\beta_{2}(t)=1,$
so $\beta_{3}(t)$ and $\beta_{4}(t)$ are given by%
\begin{equation}
\beta_{3}(t) =g\left(t\right)  +ik\left(  t\right), \label{B3}\\
\end{equation}
\begin{equation}
\beta_{4}(t)=s\left(t\right)+iw\left(t\right),  \label{B4}%
\end{equation}
where\\ 
$g\left(t\right)=-\int\frac{dt}{m\left(  t\right)  }$,   $k\left(t\right)  =2 {\displaystyle\int} f\left(t\right)dt$,\quad 
$s\left(t\right)=-\int f\left(t\right)k\left(t\right)dt$ \quad and \quad $w\left(t\right)=\int f\left(  t\right)g\left( t\right)dt$.

Substituting Eqs. $\left(  \ref{B3}\right)  $ and $\left(  \ref{B4}\right)  $
in Eq. $\left(  \ref{pI1}\right)  $ we found
\begin{equation}
I_{1}^{ph}\left(  t\right)  =p^{2}+x+\left[  g\left(  t\right)  +ik\left(
t\right) \right] p+s\left(t\right)+iw\left(t\right). \label{I_ph1}%
\end{equation}

Its eigenvalue equation is as follows
\begin{equation}
I_{1}^{ph}\left(  t\right)  \left\vert \psi\left(t\right)  \right\rangle
=\lambda_{1}\left\vert \psi\left(t\right)  \right\rangle ,
\end{equation}
in order to show that the spectrum of $I_{1}^{ph}\left(  t\right)$ is real, we
search for a metric operator that fulfills the pseudo hermiticity relation%
\begin{equation}
I_{1}^{ph\dag}\left(  t\right)  =\eta_{1}\left(  t\right)  I_{1}^{ph}\left(
t\right)  \eta_{1}^{-1}\left(  t\right)  .\label{108}%
\end{equation}
and we make the following choice for metric
\begin{equation}
\eta_{1}\left(  t\right)  =\exp[-\alpha\left(  t\right)  x-\beta\left(
t\right)  p],
\end{equation}
where $\alpha(t)$ and $\beta(t)$ are chosen as real functions in order that the metric operator $\eta_1(t)$ is Hermitian.

The position and momentum operators transform according to the transformation
$\eta_{1}\left(  t\right)  $ as
\begin{align}
\eta_{1}\left(  t\right)  x\eta_{1}^{-1}\left(  t\right) &  =x+i\beta\left(
t\right)  ,\\
\eta_{1}\left(  t\right)  p\eta_{1}^{-1}\left(  t\right) &  =p-i\alpha
\left(  t\right)  ,
\end{align}
incorporating these relationships into Eq. $\left(\ref{108}\right)  $, we
found%
\begin{align}
\alpha\left(  t\right)   &  =k\left(  t\right)  ,\\
\beta\left(  t\right)   &  =g\left(  t\right)  k\left(  t\right)  -2w\left(
t\right)  ,\text{\ }%
\end{align}
then the TD metric operator $\eta_{1}\left(  t\right)  $ is given by%
\begin{equation}
\eta_{1}\left(  t\right)  =\exp[-k\left(  t\right)  x-(g\left(  t\right)
k\left(  t\right)  -2w\left(  t\right)  )p],\label{eta1}%
\end{equation}
according to the relation $\eta_{1}\left(t\right)  =$ $\rho_{1} ^{\dag}\left(
t\right) \rho_{1}\left(  t\right)$, and since $\rho_{1}\left(t\right)$ is not unique, we can take it as a hermitian operator in order to simplify the calculations
\begin{equation}
\rho_{1}\left(  t\right)  =\exp\left[  -\frac{k\left(  t\right)  }{2}x-\left[
\frac{g\left(  t\right)  k\left(  t\right)  }{2}-w\left(  t\right)  \right]
p\right]  ,\label{rho1}%
\end{equation}
the hermitian invariant $I_{1}^{h}\left(  t\right)  $ associated with the
pseudo-hermitian invariant $I_{1}^{ph}\left(  t\right)  $ is given by
\begin{equation}
I_{1}^{h}\left(  t\right)  =\rho\left(  t\right)  I_{1}^{ph}\rho^{-1}\left(
t\right)  =p^{2}+x+g\left(  t\right)  p+\frac{k^{2}\left(  t\right)  }%
{4}+s\left(  t\right)  .\label{Ih1}%
\end{equation}

For the region $x\leq0$, we take the non-hermitian invariant $I_{2}^{ph}$ as%
\begin{equation}
I_{2}^{ph}\left(  t\right)  =\alpha_{1}\left(  t\right)  p^{2}+\alpha
_{2}\left(  t\right)  x+\alpha_{3}\left(  t\right)  p+\alpha_{4}\left(
t\right)  ,\label{216}%
\end{equation}
where $\alpha_{i}\left(t\right)$ are arbitrary complex functions to be
determined.\\ 
In the same way as the precedent case, inserting the expressions $(\ref{100'})$ and $(\ref{216})$ in
Eq. $\left(\ref{NINV}\right)$, where we take $\alpha_{1}(t)=1$ and $\alpha
_{2}(t)=-1,$ so $\alpha_{3}(t)$ and $\alpha_{4}(t)$ are given by%
\begin{align}
\alpha_{3}(t)  &  =-g\left(  t\right)  -ik\left(  t\right)  ,\text{
\ }\label{218}\\
\alpha_{4}(t)  &  =s\left(  t\right)  +iw\left(  t\right). \label{218a}%
\end{align}
Then, the final results of $I_{2}^{ph}\left(t\right)$ and $\eta_{2}\left(t\right)$ are 
\begin{equation}
I_{2}^{ph}\left(  t\right)  =p^{2}-x-\left[  g\left(  t\right)  +ik\left(
t\right)  \right]  p+s\left(  t\right)  +iw\left(  t\right),  \label{219}%
\end{equation}
\begin{equation}
\eta_{2}\left(  t\right)  =\exp[{{k\left(  t\right)  x-\left[  2w\left(
t\right)  -g\left(  t\right)  k\left(  t\right)  \right]  p}}]. \label{227}%
\end{equation}
We take $\rho_{2}\left(t\right)$ as a hermitian operator, then $\eta_{2}\left(  t\right)=\rho_{2}^{2}$,%
\begin{equation}
\rho_{2}\left(  t\right)  =\exp\left[ \frac{k\left(  t\right)  }{2}x+\left[
\frac{k\left(  t\right)  g\left(  t\right)  }{2}-w\left(  t\right)  \right]p\right]  ,\label{228}%
\end{equation}
and the related hermitian invariant $I_{2}^{h}\left(  t\right)  $ is
\begin{equation}
I_{2}^{h}\left(  t\right)  =p^{2}-x-g\left(  t\right)  p+\frac{k^{2}\left(
t\right)  }{4}+s\left(  t\right)  .\label{229}%
\end{equation}

To derive the eigenvalues equations of the invariants $I_{j}^{h}\left(
t\right) $ for the two regions $(j=1,2)$, we introduce the unitary transformations
$U_{j}(t)$
\begin{equation}
\left\vert \phi_{n,j}\left(t\right)  \right\rangle =U_{j}(t)\left\vert
\varphi_{n}  \right\rangle ,\text{ \ \ }j=1,2\label{118'}%
\end{equation}
where $\varphi_{n} $ will be determined later and
\begin{equation}
U_{1}(t)=\exp\left[  -i\frac{g\left(  t\right)  }{2}x+\frac{i}{4}\left[
k^{2}\left(  t\right)  -g^{2}\left(  t\right)  +4s\left(  t\right)  \right]
p\right]  ,\label{118}%
\end{equation}%
\begin{equation}
U_{2}(t)=\exp\left[  i\frac{g\left(  t\right)  }{2}x-\frac{i}{4}\left[
k^{2}\left(  t\right)  -g^{2}\left(  t\right)  +4s\left(  t\right)  \right]
p\right].
\end{equation}

According to these transformations, the invariants $I_{1}^{h}\left(  t\right)
$ and $I_{2}^{h}\left(  t\right)  $ turn into
\begin{align}
I_{1} &  =U_{1}^{\dag}(t)I_{1}^{h}\left(  t\right)  U_{1}(t)=p^{2}%
+x,\label{118+}\\
I_{2} &  =U_{2}^{\dag}(t)I_{2}^{h}\left(  t\right)  U_{2}(t)=p^{2}%
-x,\label{0118-}%
\end{align}
and they can be written in the following combined form
\begin{equation}
I=p^{2}+\left\vert x\right\vert .
\end{equation}
 We note here that $I$ can be considered as the Hamiltonian of a particle of mass $m_{0}=1/2$ confined in the linear symmetric potential well $\left\vert x\right\vert$. Therefore, the
eigenvalue equation of the invariant $I$
\begin{equation}
\left[  \frac{d^{2}}{dx^{2}}+\left(  \lambda_{n}-\left\vert x\right\vert
\right)  \right]  \varphi_{n}\left(  x\right)  =0,\label{eve}%
\end{equation}
is a well-known problem in quantum mechanics. The bound states $\varphi
_{n}\left(x\right)$ are given in terms of the Airy functions $Ai$ and $Bi$ \cite{Schwinger,Vallee}
\begin{equation}
\varphi_{n}\left(x\right)  =N_{n}\text{ }Ai\left(  \left\vert x\right\vert
-\lambda_{n}\right)+ N_{n}^{'}\text{ }Bi\left(  \left\vert x\right\vert
-\lambda_{n}\right), \label{ISOL1}%
\end{equation}
this solution is not relevant because $Bi\left(  \left\vert x\right\vert-\lambda_{n}\right) $ tends to infinity for $\left(\left\vert x\right\vert-\lambda_{n}\right)>0$. Thus, we take $N_{n}^{'}=0$ and the above solution reduces to
\begin{equation}
\varphi_{n}\left(  x\right)  =N_{n}\text{ }Ai\left(  \left\vert x\right\vert
-\lambda_{n}\right). \label{ISOL}%
\end{equation}
The eingenvalues $\lambda_{n}$ are determined by matching the functions 
$\varphi_{n}\left(x\right)$ and their derivatives in the two regions at the point $x=0$
\begin{align}
\varphi_{n}^{(1)}\left(  0\right)   &  =\varphi_{n}^{(2)}\left(  0\right)  ,\\
\varphi_{n}^{^{\prime}(1)}\left(0\right)   &  =\pm\varphi_{n}^{^{\prime}%
(2)}\left(  0\right) ,
\end{align}
from which there are two possibilities for $\lambda_{n}$ and the normalisation constant $N_{n}$ depending on whether $n$ is even or odd :

$\bullet$ If $n$ is even:
\begin{equation}
\lambda_{n}=-a_{\frac{n}{2}+1}^{\prime},
\end{equation}
where $a_{k}^{\prime}$ is the $k^{th}$ zero of the derivative $Ai^{\prime}$ of
the Airy function, and all values of $a_{k}^{\prime}$ are negative numbers \cite{nist}.

The normalisation constant is
\begin{equation}
N_{n}=\frac{1}{\sqrt{-2a_{\frac{n}{2}+1}^{\prime}}Ai\left(  a_{\frac{n}{2}%
+1}^{\prime}\right)  },
\end{equation}
and the corresponding eigenfunction of $I$ is
\begin{equation}
\varphi_{n}\left(x\right)  =\frac{1}{\sqrt{-2a_{\frac{n}{2}+1}^{\prime}%
}Ai\left(  a_{\frac{n}{2}+1}^{\prime}\right)  }Ai\left(  \left\vert
x\right\vert +a_{\frac{n}{2}+1}^{\prime}\right)  ,\label{fiev}%
\end{equation}

$\bullet$ If $n$ is odd:
\begin{equation}
\lambda_{n}=-a_{\frac{n+1}{2}},
\end{equation}
where $a_{k}$ is the $k^{th}$ zero of the Airy function $Ai$, and all values of $a_{k}$ are negative numbers \cite{nist}.

The normalisation constant is
\begin{equation}
N_{n}=\frac{1}{\sqrt{2}Ai^{\prime}\left(  a_{\frac{n+1}{2}}\right)  },
\end{equation}
and the corresponding eigenfunction of $I$ is
\begin{equation}
\varphi_{n}\left(  x\right)  =sgn\left(  x\right)  \frac{1}{\sqrt{2}%
Ai^{\prime}\left(  a_{\frac{n+1}{2}}\right)  }Ai\left(  \left\vert
x\right\vert +a_{\frac{n+1}{2}}\right)  .\label{fiod}%
\end{equation}

The eigenfunctions of the hermitian invariants $I_{j}^{h}\left(  t\right)  $
are written for each region as
\begin{equation}
\left\vert \phi_{n,j}\left(t\right)  \right\rangle =U_{j}(t)\left\vert
\varphi_{n}  \right\rangle,\label{fi}%
\end{equation}
then, the eigenfunctions of the pseudo-hermitian invariants $I_{j}^{ph}\left(
t\right)  $ are given by
\begin{equation}
\left\vert \psi_{n,j}\left(t\right)  \right\rangle =\rho_{j} ^{-1}\left(
t\right)U_{j}(t)\left\vert \varphi_{n}  \right\rangle
,\label{psi}%
\end{equation}
thus, the solutions of the time-dependent Schr\"{o}dinger equation $\left(
\ref{100}\right)  $ take the form%
\begin{equation}
\left\vert \Psi_{n,j}\left(t\right)  \right\rangle =e^{i\epsilon_{n}%
^{j}\left(  t\right)  }\left\vert \psi_{n,j}\left(t\right)  \right\rangle
\label{PSI}%
\end{equation}
where $\epsilon_{n}^{j}\left(  t\right)  $ is the phase ($\epsilon_{n}%
^{1}\left(  t\right)  $ for $x\geq0$ and $\epsilon_{n}^{2}\left(  t\right)  $
for $x\leq0$)$,$ which is obtained from the following relation%
\begin{align}
\dot{\epsilon}_{n}^{j}\left(  t\right)   &  =\left\langle \psi_{n,j}\left(t\right)  \right\vert \eta_{j}(t)\left[  i\frac{\partial}{\partial t}-H\left(  t\right)  \right]  \left\vert \psi_{n,j}\left(t\right)\right\rangle \nonumber\\
 & =\left\langle \phi_{n,j}\left(t\right)  \right\vert i\rho_{j}\left(  t\right)  \dot{\rho}_{j}^{-1}\left(  t\right) \left\vert \phi_{n,j}\left(t\right) \right\rangle \nonumber\\
 & - \left\langle \phi_{n,j}\left(t\right) \right\vert \rho_{j}\left(t\right)  H\left(  t\right)\rho_{j}^{-1}\left(t\right) \left\vert \phi_{n,j}\left(t\right) \right\rangle  \nonumber\\
 &+ \left\langle \phi_{n,j}\left(t\right) \right\vert  i\frac{\partial}{\partial t} \left\vert \phi_{n,j}\left(t\right) \right\rangle \nonumber\\
 & =\theta\left(t\right) -\left\langle \phi_{n,j}\left(t\right) \right\vert \frac{p^{2}}{2m\left(  t\right)  }\left\vert \phi_{n,j}\left(t\right) \right\rangle \nonumber\\
&+\left\langle \phi_{n,j}\left(t\right)\right\vert i\frac{\partial}{\partial t} \left\vert \phi_{n,j}\left(t\right) \right\rangle ,
\end{align}
where
\begin{equation}
\theta\left(  t\right)  =\frac{1}{2}f\left(  t\right)  \left[  \frac{k\left(
t\right)  }{2}g\left(  t\right)  -w\left(  t\right)  \right].
\end{equation}
Using the unitary transformations $U_{j}\left(  t\right)$, we found%
\begin{equation}
\dot{\epsilon}_{n}^{j}\left(  t\right)  =\chi^{j}\left(  t\right)  -\frac
{1}{2m\left(  t\right)  }\left\langle \varphi_{n}\left(t\right)  \right\vert
(p^{2}\pm x)\left\vert \varphi_{n}\left(t\right)  \right\rangle ,
\end{equation}
where%
\begin{equation}
\chi^{1}\left(  t\right)  =\theta\left(  t\right)  -\frac{1}{16m\left(
t\right)  }\left[  k^{2}\left(  t\right)  +3g^{2}\left(  t\right)  +4s\left(
t\right)  \right]  ,\label{xi}%
\end{equation}%
\begin{equation}
\chi^{2}\left(  t\right)  =\theta\left(  t\right)  +\frac{1}{16m\left(
t\right)  }\left[  k^{2}\left(  t\right)  -g^{2}\left(  t\right)  +4s\left(
t\right)  \right]  .
\end{equation}

From the eigenvalue equation of the invariant $I$, we have%
\begin{equation}
\left(  p^{2}\pm x\right)  \left\vert \varphi_{n}\left(t\right)
\right\rangle =\lambda_{n}\left\vert \varphi_{n}\left(t\right)
\right\rangle,
\end{equation}
then, the phases $\epsilon_{n}^{j}\left(  t\right)  $ take the form%
\begin{equation}
\epsilon_{n}^{j}\left(  t\right)  =\int\left(  \chi^{j}\left(  t\right)
-\frac{\lambda_{n}}{2m\left(  t\right)  }\right)  dt,
\end{equation}
and the solution of the TD Schr\"{o}dinger equation $\left(  \ref{100}\right)
$ is given by%
\begin{equation}
\left\vert \Psi_{n,j}\left(t\right)  \right\rangle =\exp\left[
i\epsilon_{n}^{j}\left(  t\right)  \right]  \rho_{j}\left(  t\right)
^{-1}\left\vert \phi_{n,j}\left(t\right)  \right\rangle .\label{GEN Sol}%
\end{equation}

In position representation we have%
\begin{align}
\left\langle x\left\vert \rho_{j}^{-1}\left(  t\right)  \right\vert \phi
_{j}\left(t\right)  \right\rangle &=\exp\left[i\zeta\left(  t\right)
\right]  \exp\left[  \pm\frac{k\left(  t\right)  }{2}x\right]  \nonumber\\ 
&\times \phi_{j}\left(x\pm i\left(  \frac{g\left(  t\right)  k\left(  t\right)  }{2}-w\left(
t\right)  \right)  ,t\right)  ,
\end{align}
where $\left(+\right)$ is for the positive region while $\left(-\right)$ is for the negative region, and
\begin{equation}
\zeta\left(  t\right)  =-\frac{k}{4}\left(  \frac{g\left(  t\right)  k\left(
t\right)  }{2}-w\left(  t\right)  \right)  .
\end{equation}

Then, the solution of the Schr\"{o}dinger equation for each region $\left(
\ref{GEN Sol}\right)$ can be written as
\begin{align}
\Psi_{n,j}\left(x,t\right)   &=\exp\left[i\left(
\epsilon_{n}^{j}\left(  t\right)  +\zeta\left(  t\right)  \right)  \right]
\exp\left[\pm\frac{k\left(  t\right)  }{2}x\right]  \nonumber\\ 
&\times \phi
_{n,j}\left(  x\pm i\left(  \frac{g\left(  t\right)  k\left(  t\right)  }%
{2}-w\left(t\right)\right),t\right) ,
\end{align}
and the general solution of the Schr\"{o}dinger equation $\left(\ref{100}%
\right)  $ is given by%
\begin{equation}
\Psi\left(  x,t\right) =\left\{
\begin{array}
[c]{c}%
\Psi_{n,1}\left( x,t\right)  \text{\ \ for\ } x\geq0,\\
\Psi_{n,2}\left(x,t\right)   \text{\ \ for\ } x\leq0.
\end{array}
\right.  
\end{equation}

 According to the Eqs. $\left(  \ref{ISOL}\right)  ,$ $\left(  \ref{fi}\right)
,$ $\left(  \ref{psi}\right)  $ and $\left(  \ref{PSI}\right)  $, the
probability density function is given by%
\begin{equation}
\left\vert \rho_{1}\left(  t\right)  \Psi_{n,1}\right\vert
^{2}+\left\vert \rho_{2}\left(  t\right)  \Psi_{n,2}\right\vert ^{2}   =\left\vert \phi_{n,1}  \right\vert^{2} +\left\vert \phi_{n,2}  \right\vert ^{2}  =\left\vert \varphi_{n} \right\vert ^{2},
\end{equation}
and because $\varphi_{n}\left(  x\right)  $ is determined in terms of Airy
function $Ai\left(  x\right),$ which is a real function, and according to
Eqs. $\left(  \ref{fiev}\right)  $ and $\left(  \ref{fiod}\right)$, the
probability density expression can be written as 

$\bullet$ For n is even%
\begin{equation}
\left\vert \varphi_{n}\left(x\right)\right\vert^{2}=\frac{1}{(-2a_{\frac{n}{2}+1}^{\prime
})\left [Ai(a_{\frac{n}{2}+1})\right]^{2}} \left[Ai\left(\left\vert x\right\vert +a_{\frac{n}{2}+1}^{\prime}\right)\right]^{2}, \label{densityeven}
\end{equation}
and which is represented in figure 1 for the first three even states $\left(n=0,2,4\right)$.
\begin{figure*}
\centering
{\includegraphics[width=0.8\linewidth]{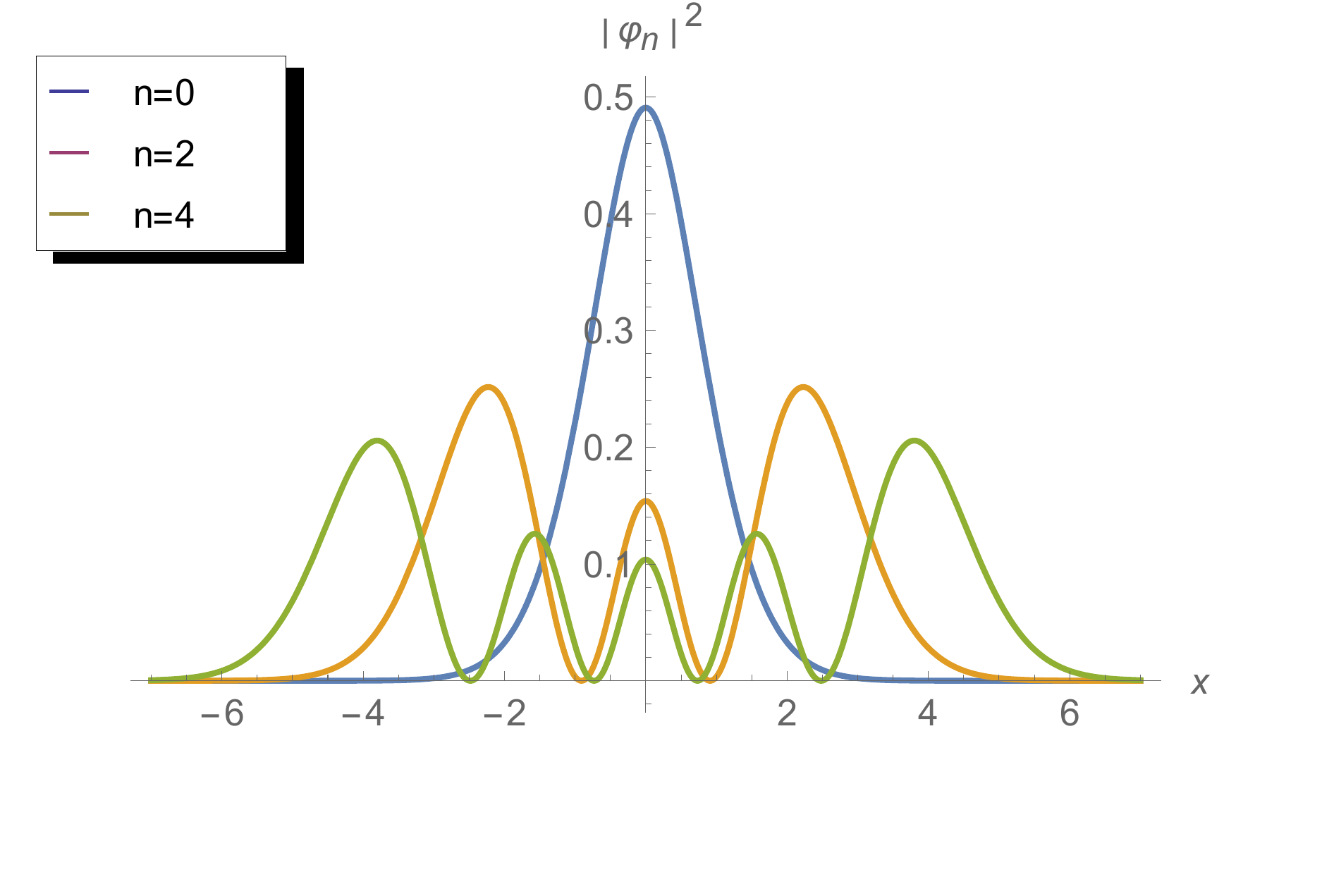}} %\hfill
\caption{Probability density of Eq. (\ref{densityeven}) for even values of $n=0,2,4$.}%
\label{fig1}%
\end{figure*}

$\bullet$ For n is odd%
\begin{equation}
\left\vert \varphi_{n}\left(x\right)\right\vert^{2}=\frac{1}{2 \left [Ai^{\prime}(a_{\frac
{n+1}{2}})\right]^{2}} \left[ Ai\left(\left\vert x\right\vert +a_{\frac{n+1}{2}%
}\right) \right]^{2}, \label{densityodd}
\end{equation}
and which is represented in figure 2 for the first three odd states $\left(n=1,3,5\right)$.
\begin{figure*}
\centering
{\includegraphics[width=0.8\linewidth]{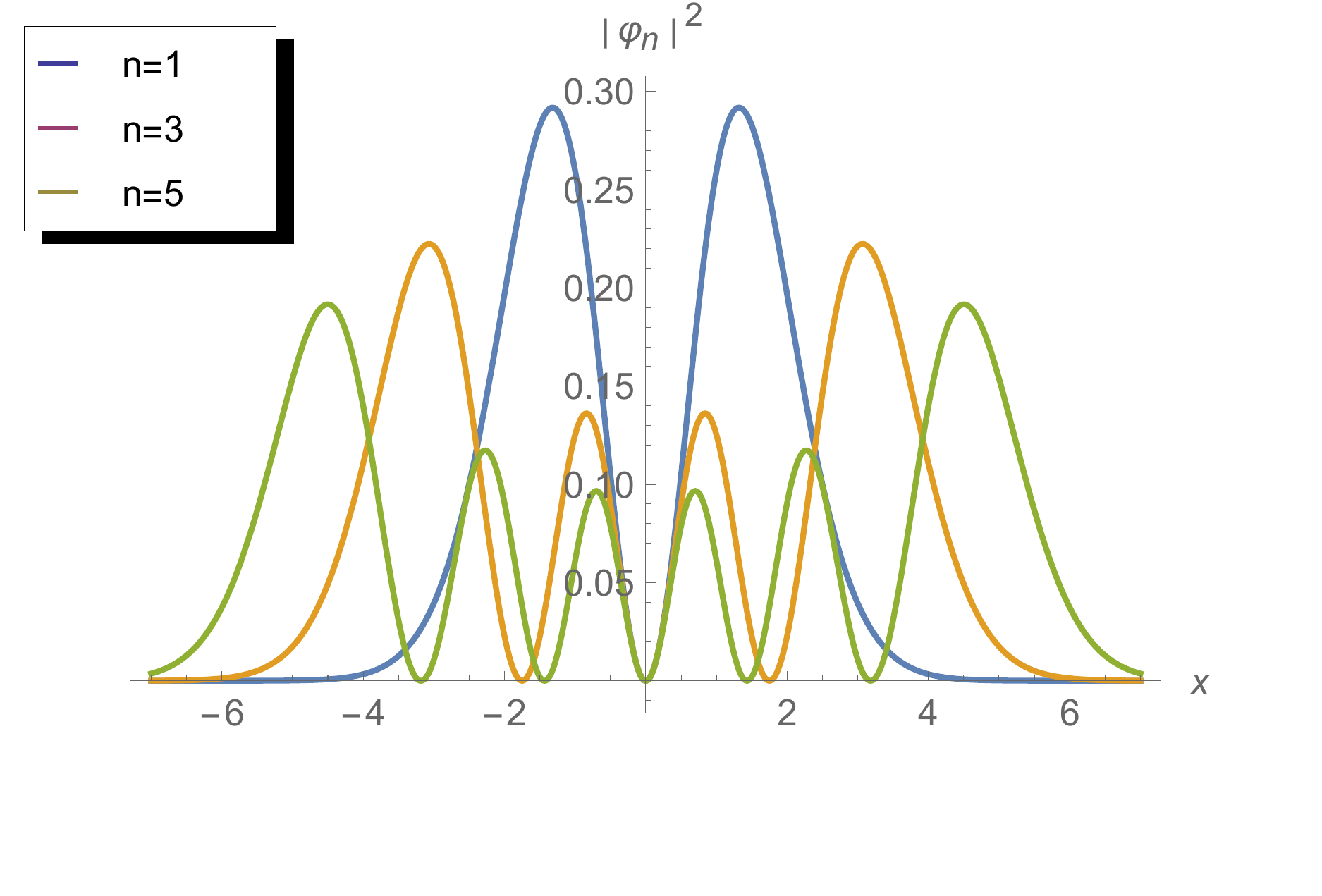}} %\hfill
\caption{Probability density of Eq. (\ref{densityodd}) for odd values of $n=1,2,3$.}%
\end{figure*} 

We note here that the probability in the region $x\leq0$ is

\begin{align}
\left\langle \Psi_{n,2}\left(t\right)  \right\vert \eta_{2}\left(  t\right)
\left\vert \Psi_{n,2}\left(t\right)  \right\rangle  & =\left\langle
\varphi_{n} \right\vert \left.  \varphi_{n} \right\rangle _{x\leq0}\nonumber\\
& =\int\limits_{-\infty}^{0}\varphi_{n}^{\ast}\left(x\right)  \varphi
_{n}\left(  x\right)  dx=\frac{1}{2},
\end{align}
and the probability in the region $x\geq0$ is 
\begin{align}
\left\langle \Psi_{n,1}\left( t\right)  \right\vert \eta_{1}\left(  t\right)
\left\vert \Psi_{n,1}\left(t\right)  \right\rangle  & =\left\langle
\varphi_{n}  \right\vert \left.  \varphi_{n}  \right\rangle _{x\geq0}\nonumber\\
& =\int\limits_{0}^{\infty}\varphi_{n}^{\ast}\left(  x\right)  \varphi
_{n}\left(  x\right)  dx=\frac{1}{2}.
\end{align}

So the two regions are equiprobable and the probability in all space is equal to one
\begin{align}
\left\langle \Psi\left(t\right)  ,\Psi\left(t\right)  \right\rangle
_{\eta}  & =\left\langle \Psi_{n,1}  \right\vert \eta
_{1}\left(  t\right)  \left\vert \Psi_{n,1}\right\rangle
+\left\langle \Psi_{n,2} \right\vert \eta_{n,2}\left(
t\right)  \left\vert \Psi_{2}\right\rangle \nonumber\\
& =\int\limits_{-\infty}^{\infty}\varphi_{n}^{\ast}\left(  x\right)
\varphi_{n}\left(  x\right)  dx=1.
\end{align}

\section{Conclusion}

The pseudo-invariant method has been used to obtain the exact analytical
solutions of the time-dependent Schr\"{o}dinger equation for a particle with
time-dependent mass moving in a complex time-dependent symmetric potential
well. We have shown that the problem can be reduced to solve a well-known
eigenvalue equation for a time-independent hermitian invariant. In fact, with
a specific choice of the TD metric operators, $\eta_{1}\left( t\right) $ and
$\eta_{2}\left( t\right) $, and the Dyson maps, $\rho_{1}\left(  t\right) $
and $\rho_{2}\left( t\right) $, and using unitary transformations, the
pseudo-invariants operators ($I_{1}^{ph}\left( t\right) $ for $x\geqslant0$
and $I_{2}^{ph}\left( t\right) $ for $x\leqslant0$) are mapped to two
time-independent Hermitian invariants $I_{1}^{h}\left( t\right) $ and
$I_{2}^{h}\left( t\right) $, which can be combined in a unique form 
$I=p^{2}+\left\vert x\right\vert $.  The latter can be considered as the
Hamiltonian of a particle confined in a linear time-independent
symmetric potential well, where its eigenfunctions are given in terms of the
Airy function $Ai$. The phases have been calculated for the two regions and
are real. Thus, the exact analytical solution of the problem has been
deduced. Finally, let us highlight the fact that the probability density 
associated with the model in question is time-independent.

%\bibliographystyle{actapoly}
%\bibliography{References.bib}

\end{document}